# In The Field Monitoring of Interactive Applications

Oscar Cornejo, Daniela Briola, Daniela Micucci, Leonardo Mariani
Department of Informatics, Systems and Communication
University of Milano Bicocca, Milan, Italy
{oscar.cornejo, daniela.briola, micucci, mariani}@disco.unimib.it

*Abstract*—Monitoring techniques can extract accurate data about the behavior of software systems. When used in the field, they can reveal how applications behave in real-world contexts and how programs are actually exercised by their users. Nevertheless, since monitoring might need significant storage and computational resources, it may interfere with users activities degrading the quality of the user experience.

While the impact of monitoring has been typically studied by measuring the overhead that it may introduce in a monitored application, there is little knowledge about how monitoring solutions may actually impact on the user experience and to what extent users may recognize their presence.

In this paper, we present our investigation on how collecting data in the field may impact the quality of the user experience. Our initial results show that non-trivial overhead can be tolerated by users, depending on the kind of activity that is performed. This opens interesting opportunities for research in monitoring solutions, which could be designed to opportunistically collect data considering the kind of activities performed by the users.

*Keywords*-Monitoring, dynamic analysis, user experience.

## I. Introduction

*In-the-field executions* might be an invaluable source of information for complementing and completing the knowledge gained with in-house testing and analysis. For instance, in-the-field executions may indicate how end-users use applications, how applications interact with several diverse environments, what adverse conditions may trigger failures, and what behaviors are more reliable than others.

Indeed in-the-field executions have been already exploited as source of information to assist testing and analysis activities. For example, crash reporting functionalities are extensively present in commercial and open source software [1], [2], [3]. In general there are a number of tasks that can benefit from data collected from the field. For example, bug isolation techniques can greatly benefit from the volume of data that can be automatically extracted from the field [4], [5]. Field data may also help profiling applications [6], improving code coverage [7], [8], discovering and reproducing failures [9], [10], and controlling software evolution [11].

Although field executions can be relevant and useful, and can enable a range of interesting analyses, collecting data beyond simple crash reports could be *extremely expensive and challenging* [12]. In fact, while crash reporting simply requires taking a snapshot of the system at the time of the crash, other solutions require monitoring applications more extensively, potentially affecting the quality of the user experience. For instance, collecting the sequences of function calls produced by a monitored system may slow down every interaction with the system, deteriorating the user experience.

When the events to be collected are independent, techniques such as probabilistic monitoring [4], [5], which samples executions with a given probability, and distributive monitoring [11], [13], which distributes the monitoring workload among multiple machines running the same application, might be exploited to reduce the overhead. Unfortunately, the useful information that can be collected from the field is seldom in the form of independent events [9], [12], [6] and monitoring sequences of events can be extremely expensive.

The impact of extensive monitoring on the *user experience* has not been studied yet. Data about the overhead are useful, but represent a partial information that does not fully capture the effect of monitoring. For instance, whether an overhead of 20% is acceptable or not depends on the way it affects the user experience, and it is hard to tell a priori. For example, increasing by 20% the time that every menu item requires to open may introduce a small but annoying slowdown to actions that should be instantaneous from a user perspective. On the contrary, taking 20% more time on the execution of a query might be acceptable for users, as long as the total time does not exceed their expectations. It is thus important to investigate the relation between the overhead introduced by monitoring techniques and the user experience, to understand how to seamlessly and feasibly collect data in the field.

In this paper, we report our initial experience with the investigation of the relation between the overhead and the quality of the experience as perceived by the users. Although results are preliminary, they already bring interesting insights into the problem of monitoring interactive applications in the field, compared to studies that consider the overhead neglecting its acceptability for the end users [6]. In particular, our experience about collecting function call sequences from seven popular applications produced three key findings.

*Non trivial overhead can be tolerated in the field*: contrarily to the common belief that a small overhead might be difficult to tolerate in the field, we found that an overhead up to 30% can be hardly recognized by users. Although systematically slowing down the system by 30% might still be problematic, sporadically introducing a non-trivial overhead might be feasible. This result opens to the interesting opportunity of embedding non-trivial monitoring and analysis solutions into the software running in the field.

*Monitoring strategies should adapt to the running tasks*: a same monitoring strategy (e.g., collecting function call sequences) does not introduce the same overhead over all the functionalities of an application, but the overhead is distributed unevenly across them. This calls for solutions that are aware of the status of the system and adjust their behavior dynamically to prevent slowdowns that can be recognized by the users.

*Collecting data during simple computations is less intrusive than collecting data during complex computations:* not all the functionalities respond to the overhead in a same way. In particular, simple functionalities tend to tolerate the overhead better than complex functionalities. For instance, collecting data while users navigate the GUI is less likely to affect the user experience compared to collecting data while users open or save files. This calls again for solutions that can dynamically control the amount of collected data to prevent any impact on the user experience.

This paper is organized as follows. Section II describes our experimental setup. Section III reports and interprets the results. Section IV provides final remarks.

## II. EXPERIMENT DESIGN

The objective of our experiment is to answer to the research question: *"How does collecting field data affect the user experience?"*. In this initial study, we investigated this research question in a restricted, although common, scenario, that is, recording the sequence of function calls produced by an application. Many techniques exploit this type of field data, such as techniques for reproducing failures [9], profiling users [6], and controlling software evolution [11].

To address this general question, we identified three specific research questions to be investigated:

**RQ1 - What is the correlation between the overhead and the user experience?** This research question analyzes the relation between the overhead produced by a monitoring activity and its impact on the user experience.

**RQ2 - Is the effectiveness of monitoring dependent on memory consumption?** This research question investigates how the resources allocated for monitoring impact on the effectiveness of data collection.

**RQ3 - Is the observed overhead dependent on the type of application?** This research question investigates the relation between the characteristics of the monitored application and the observed overhead.

To investigate these research questions, we used the following procedure. We selected seven widely used programs of different sizes and complexity: MS Excel 2016, MS Outlook 2016, Notepad++ 6.9.2, Paint.NET 4.0.12, Winzip 20.5, and Adobe Reader DC 2015. To collect sequences of function calls from these applications, we instrumented the software using a probe that we implemented with the Intel Pin Binary Instrumentation tool[1]. The probe can be configured to use buffers of different sizes to store data in memory before flushing data into a file.

[1]https://software.intel.com/en-us/articles/pin-a-dynamic-binary-instrumentation-tool

To run each application, we have implemented a *Sikuli*[2] test case that can be automatically executed to cover a typical usage scenario. Each test case includes from 11 to 32 user actions, with a mean of 16 user actions per test. To assess the impact of the monitor, we measured the *overhead* and we estimated its effect on the *user experience*. To accurately investigate both factors, we collected data at the granularity of the individual actions performed in the tests. That is, if a test case executes actions $a_1 \ldots a_n$, we collect data about the overhead and its impact on the user experience for each action $a_i$ with $i = 1 \ldots n$. We collected data for both the application without the probe and the application instrumented with our probe configured with buffers of different sizes: 0MB (data is immediately flushed to disk), 1MB, 25MB, 50MB, 75MB, 100MB, and 200MB. Experiments have been executed on a Window 7 - 32bit machine equipped with 4GB of RAM. Each test has been repeated 5 times and mean values have been used to mitigate any effect due to the non-determinism of the execution environment. Overall, we collected near 4,000 samples about the execution time of the actions.

While the overhead can be measured as the additional time consumed by an application due to the presence of the monitor, it is important to discuss how we estimated the effect of the monitor on the user experience. In principle, assessing if a given overhead may or may not annoy users requires direct user involvement. However, user studies are expensive and can be hardly designed to cope with a volume of samples like the ones that we collected. We thus relied on the results, produced with studies based on actual users and physical measurements, already available from the human-computer interaction domain. In particular, we used the well-known and widely accepted classification proposed by Seow [14] of the System Response Time (the time taken by an application to answer to a user request) that can be associated with each action based on its nature. In this classification, actions are organized in four categories.

- *Instantaneous*: these are the most simple actions that can be performed on an application, such as entering inputs or navigating throughout menus. Users expect to receive an answer by $100 - 200$ms at most.
- *Immediate*: these are actions that are expected to generate acknowledgments or very simple outputs. Users expect to receive an answer by $0.5 - 1$s at most.
- *Continuous*: these are actions performing operations that are requested to produce results within a short time frame to not interrupt the dialog with the user. These functions are expected to produce an answer in $2 - 5$s at most, depending on the complexity of the operation that is executed. We assume that a simple continuous action should produce an answer by $2 - 3.5$s and more complex continuous actions should produce an answer by $3.5 - 5$s.
- *Captive*: these are actions requiring some relevant processing for which users will wait for results, but will also give up if a response is not produced within a certain time. These actions are expected to produce an answer by $7-10$s.

[2]http://sikulix.com

We attribute categories to actions based on their execution time when no overhead is introduced in the system. We use the lower limit of each category to this end. For instance, actions that take at most 100ms are classified as instantaneous, while actions that take more than 100ms but less than 1s are classified as immediate.

We thus estimate the impact of the overhead on the user experience by measuring the number of *slow actions*, that is, the actions that exceed the upper limit of the response time for their category once affected by the overhead. According to this classification, we assume that the response time of an action is acceptable by users as long as it is below the upper limit of the category the action belongs to. Thus, an overhead that increases the response time of an action without exceeding the upper limit of the category (e.g., an instantaneous action that takes less than 200ms once affected by the overhead) would be hardly noticeable by users. On the contrary, if the overhead increases the response time of an action above the upper limit of the category (e.g., an instantaneous action that takes more than 200ms once affected by the overhead), the execution time of the action would likely violate the user expectation, and the slowdown would be recognizable by the users.

## III. RESULTS

In this section, we report the results that we obtained for the three research questions.

### A. Research Question 1

To answer this research question, we compute the percentage of slow actions distinguishing among instantaneous, immediate, simple, and complex continuous actions. We plot these percentages in Figure 1, considering overhead intervals that produce similar results. We also indicate the overall percentage of slow actions in each overhead interval. We do not plot the results for captive actions because we had only few actions belonging to this category in the tests, thus the collected data are insufficient to produce relevant insights. However, the captive actions have been included in the computation of the overall percentage of slow actions for each overhead interval.

We can first observe that an overhead up to 30% never caused a noticeable slowdown for any kind of action. This result is partially in contrast with the intuition that only a very small overhead could be tolerated in the field.

An overhead in the range $30-80\%$ makes only a few actions exceed the maximum response time for their category. In particular, only complex continuous actions show slowdowns that are likely to be perceived by users, while the overall percentage of slow actions is low. This suggests that simple computations can be safely monitored compared to complex operations, which require more attention.

An overhead in the range $80-180\%$ turns 35% of the actions into slow actions in average. Almost all categories of actions are affected, again with a greater potential impact on the most complex operations. Higher overhead values ($>180\%$) seriously compromise the response time of most of the actions. Note that the plot reports a significant effect on the instantaneous actions only when the overhead is greater than 600%. This is probably due to the simplicity of the actions, which are still fast even for high percentage overhead.

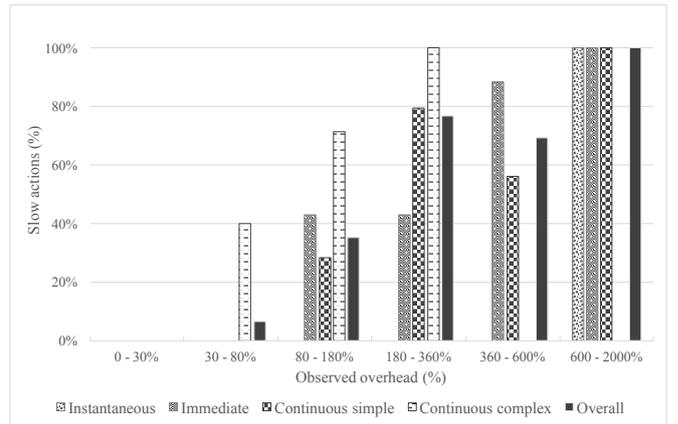

Fig. 1. Percentage of slow actions for different overhead intervals.

TABLE I
EFFECTIVENESS FOR DIFFERENT BUFFER SIZES.

| buffer size (MB) | slow actions (num (%)) | overhead (%) | RAM consumption (MB) |
|---|---|---|---|
| 0 | 20 (18%) | 170% | 192 |
| 1 | 15 (13%) | 133% | 198 |
| 25 | 16 (14%) | 104% | 243 |
| 50 | 17 (15%) | 97% | 281 |
| 75 | 21 (18%) | 128% | 321 |
| 100 | 16 (14%) | 108% | 355 |
| 200 | 16 (14%) | 93% | 485 |

In a nutshell, these results show that a non trivial overhead (e.g., up to 30%) can likely be introduced in the field with little impact on the users. This result is coherent with the study described in [15], which shows that users are usually unable to identify time variations smaller than 20%.

On the other hand, different functionalities react differently to the same amount of overhead, thus a non-intrusive monitoring technique should control the overhead per action to potentially prevent any impact on the user experience.

### B. Research Question 2

Table I shows how results change when using buffers of different sizes. We report the average number and percentage of slow actions, the average overhead introduced by the monitor, and the average amount of additional RAM consumed by the monitor, measured across all the test cases and all the applications.

The percentage of slow actions is quite stable, near 14%, regardless of the different sizes of the buffer. Thus, the specific size of the buffer is likely to have little impact on the results, as long as a buffer is used. In fact, the worst result is obtained when data is immediately written on file (buffer of size 0). For the specific set of applications that we considered, a buffer of size 50MB seems to represent a good compromise between the amount of memory consumed and the observed overhead.

TABLE II
FUNCTION CALLS RATE, OVERHEAD, AND SLOW ACTIONS.

| application | function calls rate (FCalls/s) | overhead (%) | slow actions (%) |
|---|---|---|---|
| Winzip | 2,216 | 23.7% | 7.14% |
| Paint.NET | 4,390 | 48.4% | 12.09% |
| Acrobat DC | 9,009 | 30.0% | 7.69% |
| VLC | 21,663 | 41.9% | 7.14% |
| Notepad++ | 56,101 | 12.2% | 2.86% |
| MS Excel | 2,977,154 | 387.9% | 51.95% |
| MS Outlook | 3,029,380 | 269.4% | 48.05% |

Note that an average overhead of 14% for a given configuration of the buffer does not correspond to a same overhead for every single action performed in each test case. Monitoring may have a different impact on different actions. This is confirmed by our observations. The average and maximum variance of the overhead internally to a same test case has been 373% and 3.324%, respectively. This further stresses the fact that a non-intrusive monitoring technique should control the overhead on a per-action basis, and cannot be configured once for all for an application.

*C. Research Question 3*

Finally, we investigate the presence of a relationship between the type of subject application and the overhead. We conjecture that the overhead may depend on the rate of monitored events (function calls in our case) produced by a subject application. To this end, we compute the average rate of function calls, that is, the number of functions invoked per second, for all the subject applications and for all the tests. Moreover, we compute the average overhead introduced in each subject application and the percentage of slow actions. The results are reported in Table II.

Although the overhead does not strictly grow for an increasing function call rate, we can clearly distinguish two main cases. Small applications producing a limited number of function calls, that is, below 56K calls per second, show an overhead below 50% and a rate of slow actions below 13%. Instead, large and complex applications with a function call rate close to 3 millions of calls per second show an overhead higher than 250% with 50% of the actions producing a slowdown that can be likely recognized by the users. We confirmed the correlation between the function calls rate and the overhead computing the Person correlation index, which returned a value of 0.97 ($p\text{-value} < 10^{-3}$) for the correlation between the overhead and the function calls rate, and a value of 0.99 ($p\text{-value} < 10^{-4}$) for the correlation between the number of slow actions and the function calls rate.

This result implies that in some cases the structural characteristics of an application, such as its internal design, may play a relevant role in estimating the cost of monitoring. For instance, an application organized into several functions can be more expensive to monitor than a program implemented monolithically, for probes collecting function calls sequences.

## IV. CONCLUSIONS

In this paper we investigated how the overhead introduced by monitoring probes that collect function calls may impact on the user experience. We focused our study on widely adopted interactive applications and we exploited a well-known classification about the system response time [14] to estimate the impact of monitoring on the user experience.

Our results suggest that a non-trivial overhead might be tolerated by users. This creates interesting research opportunities about the deployment of testing and analysis techniques in the field. On the other hand, a same probe does not introduce a same overhead across all the functionalities of an application. Thus, the design of non-intrusive techniques requires monitors that are aware of the current status of the system and that can opportunistically decide if and how much data to collect.

In our study we measured the impact of monitoring sequences of function calls without directly involving the users of the monitored applications. In the future, we plan to corroborate our results with ad-hoc user studies and to extend the scope of the study to the detection of other events, such as the execution of program statements.

ACKNOWLEDGMENT

This work has been partially supported by the H2020 Learn project, which has been funded under the ERC Consolidator Grant 2014 program (ERC Grant Agreement n. 646867).